\documentclass[thmsa,oneside,a4paper]{article}
\usepackage{amsfonts}
\usepackage{amsmath}
\usepackage{graphicx}

\input{tcilatex}

\begin{document}

\title{Representation Theory and Baxter's TQ equation for the six-vertex
model. A pedagogical overview.}
\author{{\Large Christian Korff\medskip } \\
\emph{School of Mathematics, University of Edinburgh}\\
\emph{Mayfield Road, Edinburgh EH9 3JZ, UK}}
\maketitle

\section{Introduction}

Recent years have seen renewed interest in the six and eight-vertex model at
rational coupling values, that is when the crossing parameter is evaluated
at roots of unity. Deguchi, Fabricius and McCoy pointed out that the extra
degeneracies in the spectrum of the six-vertex transfer matrix can be
understood in terms of an affine symmetry algebra in certain commensurate
spin-sectors \cite{DFM}. This raised the question of how the representation
theory of this algebra manifests itself in the Bethe ansatz and determines
the level of degeneracy \cite{DFM,FM01a,FM01b}. The analogous investigation
for the eight-vertex model \cite{FM8v,FM8v2,FM8v3,FMproc} has led to new
developments and a better understanding of Baxter's celebrated $TQ$ equation 
\cite{Bx72,Bx73,BxBook}. Here $T$ denotes the transfer matrix and $Q$ stands
for the auxiliary matrix. In the trigonometric limit some of the zeroes of
the auxiliary matrix correspond to the solutions of the six-vertex Bethe
ansatz equations first derived in \cite{Lieb67,St67}.

The significance of a better understanding of the six as well as the
eight-vertex model at roots of unity has been further highlighted by results
when the order of the root of unity is three. Then Baxter's $TQ$ equation
can be explicitly solved for the groundstate eigenvalue provided the square
lattice has an odd number of columns; see \cite{FSZ,S01}\ for the six and 
\cite{FM8v3,BM}\ for the eight-vertex case. One finds two linearly
independent solutions. These arise because the $TQ$ equation is a second
order difference equation, see the discussion in \cite{KWLZ} and \cite{PrSt}%
. However, both linearly independent solutions do not always exist at roots
of unity. An additional aspect which motivated further investigation is the
connection with combinatorial aspects such as the enumeration of
alternating-sign matrices or plane partitions, see e.g. \cite{Ku96,RS,BGN}.

\subsection{Auxiliary matrices for the six-vertex model}

The described developments prompted the series of papers \cite%
{KQ,KQ2,KQ3,KQ4,KQ5} on a representation theoretic construction of $Q$%
-operators for the six-vertex model at roots of unity. See also \cite{BS90}
for an earlier, but different construction and \cite{AF97,RW02} for a
discussion away from roots of unity. The idea to use representation theory
of quantum groups to solve Baxter's $TQ$ equation (or generalizations
thereof) on the level of operators parallels the approach \cite{BLZ99} put
forward in the context of the Liouville model. However, there are subtle
differences between the continuum theory and the model on the finite lattice
and the root of unity case is only marginally discussed in \cite{BLZ99}.

In particular, at roots of unity the auxiliary space used in the
construction of $Q$ can be kept finite-dimensional (this will be explained
in more detail below) which simplifies calculations drastically and prevents
any problems with convergence; see the discussions in \cite{RW02}\ and \cite%
{KQ3}. For numerical calculations this is of great practical importance.

But why using Baxter's $TQ$ equation for the six-vertex model at all instead
of the Bethe ansatz? From the brief outline given above it should be clear
that we are interested in an approach which applies to both the six and the
eight-vertex model. This rules out the Bethe ansatz which rests on the
conservation of the total spin or the existence of a\ pseudovacuum. But
there are additional equally significant reasons:

\begin{itemize}
\item To see the full symmetry present at roots of unity one should not
choose a basis of eigenvectors which diagonalizes the total spin-operator 
\emph{even for the six-vertex model}. By dropping this requirement a much
wider set of solutions to a generalization of Baxter's TQ equation can be
obtained which do not preserve the total spin and reveal a rich geometric
structure. This provides a new, different perspective on the symmetries at
roots of unity; see \cite{KQ} and the discussion in Section 4.2 of this
article.

\item The other aspect is the existence of a second linearly independent
solution to the $TQ$ equation \cite{KWLZ,PrSt}. One solution gives the Bethe
roots above the equator, the second solution, which is related to the first
by spin-reversal, yields the Bethe roots below the equator. The explicit
construction of $Q$-operators provides the platform to rigorously prove
existence of these solutions and discuss the question of completeness; see 
\cite{KQ5}. This discussion is closely related to the aforementioned
explicit ground state solutions \cite{S01,FM8v3,BM} and the connection
between the $TQ$ equation in conformal field theory and ordinary
differential equations \cite{ODEIM}.
\end{itemize}

\subsection{The construction procedure}

Having made the case of using the concept of auxiliary matrices for the
six-vertex model let us briefly comment on the choice of the construction
procedure. There are a number of technical difficulties with the
construction described in \cite{BxBook} which can be overcome by the use of
representation theory of quantum groups. The advantages are:

\begin{enumerate}
\item A simple algebraic form of the auxiliary matrix which drastically
simplifies calculations and rests on the Yang-Baxter equation with its
underlying algebraic framework. It applies to lattices with an even as well
as an odd number of columns.

\item A representation theoretic derivation of functional equations. In
particular, there is no need for an ``a priori''\ knowledge of the form of
the $TQ$ equation.

\item It enables a generalization to a wider class of models associated with
quantum groups of higher rank and is independent of the choice of the
representation chosen for the quantum space.
\end{enumerate}

\subsection{Outline of the article}

As the title already indicates this article is intended to provide an
easy-to-digest overview of the construction of auxiliary matrices and their
relation to the special symmetries at roots of unity. We will therefore omit
any technical computations and proofs referring the interested reader to the
original manuscripts \cite{KQ,KQ2,KQ3,KQ4,KQ5}. Instead we will emphasize
that solving the $TQ$ equation is a simple step-by-step procedure which we
illustrate with numerous diagrams.

Section 2 briefly reviews the definition of the six-vertex model and its
underlying quantum group structure. Section 3 explains in general and simple
terms how the $Q$-operator is constructed and the $TQ$ equation is derived.
Section 4 makes contact with the affine symmetry algebra at roots of unity
and a geometric picture of the auxiliary matrices. For a particular
subvariety of solutions to the $TQ$ equation the spectrum is analyzed.
Section 5 states some concluding remarks.

\section{Preliminaries on the six-vertex model}

In order to keep this article self-contained we repeat some well-known facts
from the definition of the six-vertex model and its associated quantum group
structure.

\subsection{Definition}

Consider an $M\times M^{\prime }$ square lattice where one assigns to each
vertex one of the six configurations depicted in Figure 1. Each
configuration occurs with a probability determined by the Boltzmann weights $%
a,b,c,c^{\prime }$.

\bigskip
\begin{center}
\includegraphics[totalheight=1.75cm]{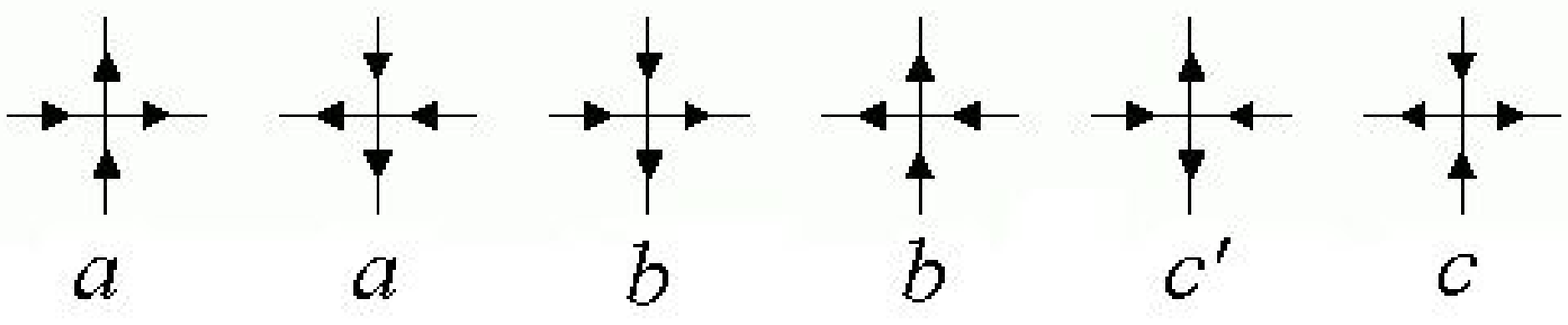}
\end{center}
\begin{center}
{\small Figure 1. The six allowed vertex configurations and their
Boltzmann weights.}
\end{center}\bigskip

The partition function $Z=\limfunc{Tr}_{\mathcal{H}}T^{M^{\prime }}$of the associated statistical
six-vertex model can be written in terms of the transfer matrix, 
\begin{equation}
T=\limfunc{Tr}_{V_{0}}R_{0M}R_{0M-1}\cdots R_{01}\in \limfunc{End}\mathcal{H}%
,\quad \mathcal{H}=(\mathbb{C}^{2})^{\otimes M}\ .  \label{T}
\end{equation}%
Here $\mathcal{H}$ is known as \textquotedblleft quantum
space\textquotedblright , $V_{0}\cong \mathbb{C}^{2}$ is called
\textquotedblleft auxiliary space\textquotedblright . The matrix $R$ is
defined over $\mathbb{C}^{2}\otimes \mathbb{C}^{2}$ and contains the
Boltzmann weights associated with the different vertex configurations, 
\begin{equation}
R=\tfrac{a+b}{2}1\otimes 1+\tfrac{a-b}{2}\,\sigma ^{z}\otimes \sigma
^{z}+c\,\sigma ^{+}\otimes \sigma ^{-}+c^{\prime }\sigma ^{-}\otimes \sigma
^{+}=\left( 
\begin{array}{cccc}
a &  &  &  \\ 
& b & c &  \\ 
& c^{\prime } & b &  \\ 
&  &  & a%
\end{array}%
\right) \,.  \label{R}
\end{equation}%
The symbols $\{\sigma ^{x},\sigma ^{y},\sigma ^{z}\}$ denote the Pauli
matrices with $\sigma ^{\pm }=(\sigma ^{x}\pm i\sigma ^{y})/2$ and the lower
indices in (\ref{T}) indicate on which pair of spaces the $R$-matrix acts in
the $(M+1)$-fold tensor product of $\mathbb{C}^{2}$. Two transfer matrices
commute with each other provided their Boltzmann weights leave the quantity $%
\Delta =\left( a^{2}+b^{2}-cc^{\prime }\right) /2ab$ invariant. The
well-known symmetries of the model are expressed in terms of the following
commutators 
\begin{equation}
\lbrack T,S^{z}]=[T,\mathfrak{R}]=[T,\mathfrak{S}]=0\;,  \label{symm}
\end{equation}%
where the respective operators are defined as 
\begin{equation}
S^{z}=\frac{1}{2}\sum_{m=1}^{M}\sigma _{m}^{z},\text{\quad }\mathfrak{R}%
=\sigma ^{x}\otimes \cdots \otimes \sigma ^{x}=\prod\limits_{m=1}^{M}\sigma
_{m}^{x},\text{\quad }\mathfrak{S}=\sigma ^{z}\otimes \cdots \otimes \sigma
^{z}=\prod\limits_{m=1}^{M}\sigma _{m}^{z}\;.  \label{SzRS}
\end{equation}%
These symmetries hold for spin-chains of even as well as odd length. We now
review the connection of this model with the affine quantum group $U_{q}(%
\widehat{sl}_{2})$. This will set the stage for the representation theoretic
construction of the auxiliary matrices.

\subsection{The quantum group $U_{q}(\widehat{sl}_{2})$}

There is a quasi-triangular Hopf algebra $U_{q}(\widehat{sl}_{2})$ which is
generated by the six elements \{$e_{i},f_{i},q^{\pm h_{i}}$\}$_{i=0,1}$
obeying the relations 
\begin{eqnarray}
q^{h_{i}}q^{h_{j}} &=&q^{h_{j}}q^{h_{i}},\quad
q^{h_{i}}q^{-h_{i}}=q^{-h_{i}}q^{h_{i}}=1,  \label{AQG} \\
q^{h_{i}}e_{j}q^{-h_{i}} &=&q^{\mathcal{A}_{ij}}e_{j},\quad
q^{h_{i}}f_{j}q^{-h_{i}}=q^{-\mathcal{A}_{ij}}f_{j}, \\
\lbrack e_{i},f_{j}] &=&\delta _{ij}~\frac{q^{h_{i}}-q^{-h_{i}}}{q-q^{-1}}%
~,\quad i,j=0,1,\quad \mathcal{A}_{ij}=(-1)^{i+j}2\ .
\end{eqnarray}%
In addition, for $i\neq j$ the $q$-deformed Chevalley-Serre relations hold, 
\begin{equation}
x_{i}^{3}x_{j}-[3]_{q}x_{i}^{2}x_{j}x_{i}+[3]_{q}x_{i}x_{j}x_{i}^{2}-x_{j}x_{i}^{3}=0,\quad x=e,f\;.
\label{CS}
\end{equation}%
Further defining relations can be found in e.g. \cite{CPbook}. For our
present purposes only one more structure is important, the coproduct of $%
U_{q}(\widehat{sl}_{2})$ which is given by\footnote{%
This choice of Hopf algebra is different from the one which often is used in
the physics literature involving a symmetric coproduct. See the comments in 
\cite{KQ}, Section 2.1, pages 5237-8. We comment on this further in Section
4.1 of this article.} 
\begin{equation}
\Delta (e_{i})=1\otimes e_{i}+q^{h_{i}}\otimes e_{i},\quad \Delta
(f_{i})=f_{i}\otimes q^{-h_{i}}+1\otimes f_{i},\quad \Delta
(q^{h_{i}})=q^{h_{i}}\otimes q^{h_{i}}\;.  \label{cop}
\end{equation}%
There is an opposite coproduct $\Delta ^{\text{op}}$ which is obtained by
permuting the two factors. Moreover, there exists an abstract universal $R$%
-matrix intertwining these two coproduct structures 
\begin{equation}
\mathbf{R\,}\Delta (x)=\Delta ^{\text{op}}(x)\,\mathbf{R},\quad x\in U_{q}(%
\widehat{sl}_{2}),\quad \mathbf{R}\in U_{q}(b_{+})\otimes U_{q}(b_{-})\;.
\label{inter}
\end{equation}%
Here $U_{q}(b_{\pm })$ denote the upper and lower Borel subalgebra generated
by $\{e_{i},q^{h_{i}},q^{-h_{i}}\}$ and $\{f_{i},q^{h_{i}},q^{-h_{i}}\}$,
respectively. In a particular representation of the quantum group the
intertwiner can now be identified with the building blocks of the transfer
matrix, i.e. the $R$-matrix (\ref{R}).

\subsection{Evaluation representation of spin $\frac{1}{2}$}

Let $z\in \mathbb{C}$ be nonzero then the mapping 
\begin{equation}
e_{0}\rightarrow z\,f,\;f_{0}\rightarrow z^{-1}e,\;q^{h_{0}}\rightarrow
q^{-h},\;e_{1}\rightarrow e,\;f_{1}\rightarrow f,\;q^{h_{1}}\rightarrow
q^{h}\quad .  \label{ev}
\end{equation}%
defines an algebra homomorphism ev$_{z}:U_{q}(\widehat{sl}_{2})\rightarrow
U_{q}(sl_{2})$. Here $U_{q}(sl_{2})$ is isomorphic to the Hopf algebra
generated by either \{$e_{1},f_{1},q^{h_{1}}$\} or \{$e_{0},f_{0},q^{h_{0}}$%
\}. Denote by $\pi _{z}=\pi \circ ev_{z}:U_{q}(\widetilde{sl}%
_{2})\rightarrow \limfunc{End}\mathbb{C}^{2}$ the spin $1/2$ evaluation
representation of the quantum group defined by the relations 
\begin{equation}
\pi (e)=\sigma ^{+},\quad \pi (f)=\sigma ^{-},\quad \pi (q^{h})=q^{\sigma
^{z}}\quad .  \label{pin}
\end{equation}%
With respect to this representation one can now consider the intertwiner of
the tensor product $\pi _{z}\otimes \pi _{1}$, i.e. 
\begin{equation}
R(z)\left( \pi _{z}\otimes \pi _{1}\right) \Delta (x)=\left[ \left( \pi
_{z}\otimes \pi _{1}\right) \Delta ^{\text{op}}(x)\right] R(z),\quad x\in
U_{q}(\widehat{sl}_{2})\ .  \label{fusL}
\end{equation}%
The matrix elements of the intertwiner can be explicitly computed using and
one finds up to an arbitrary overall normalization factor that it coincides
with the six-vertex $R$-matrix with the following parametrization of the
Boltzmann weights, 
\begin{equation*}
a=zq^{\frac{3}{2}}-q^{-\frac{1}{2}},\quad b=zq^{\frac{1}{2}}-q^{\frac{1}{2}%
},\quad c=(q-q^{-1})q^{\frac{1}{2}},\quad c^{\prime }=(q-q^{-1})zq^{\frac{1}{%
2}},\quad \Delta =\frac{q+q^{-1}}{2}\ .
\end{equation*}%
We could have chosen a different parametrization but this one ensures that
the eigenvalues of $T$ are polynomials in $z$ and that the functional
equations we are going to derive look more symmetric.

\section{The construction of $Q$}

We will now solve the eigenvalue problem of the six-vertex transfer matrix
by introducing an additional matrix $Q^{p}$ which commutes with $T$ and
satisfies a functional equation of the following type 
\begin{equation}
T(z)Q^{p}(z)=Q^{p}(z)T(z)=\phi _{1}^{M}(z)Q^{p^{\prime }}(zq^{2})+\phi
_{2}^{M}(z)Q^{p^{\prime \prime }}(zq^{-2})\ .  \label{TQ}
\end{equation}%
Here $\phi _{1},\phi _{2}$ are scalar functions depending on $z,q$ and the
upper index of the auxiliary matrix $Q^{p}$ denotes the possible dependence
on additional free parameters which we collectively denote by $p$. Note that
these parameters are allowed to shift in the functional equation. This is
one of the main differences with Baxter's procedure \cite{BxBook}. Although
it appears marginal at first sight this modification is crucial to maintain
a simple algebraic form of the auxiliary matrix. The free parameters are
also needed to break the symmetries of the transfer matrix such as
spin-reversal symmetry and, when $q$ is a root of unity, spin-conservation
as well as the affine symmetry \cite{DFM}; see \cite{KQ} for details.

\subsection{The Yang-Baxter equation}

In contrast to Baxter's approach described in Chapter 9 of his book \cite%
{BxBook} we now first ensure that the auxiliary matrix commutes with the
transfer matrix and then afterwards derive the functional equation. Also our
algebraic form of the auxiliary matrix differs from the one in \cite{BxBook}%
. Namely, to ensure the greatest possible compatibility with the definition
of the transfer matrix we choose the auxiliary matrix to be of the form 
\begin{equation}
Q^{p}(w)=\limfunc{Tr}_{V_{0}=\pi
_{w}^{p}}L_{0M}^{p}(w)L_{0(M-1)}^{p}(w)\cdots L_{01}^{p}(w)  \label{Qp}
\end{equation}%
with $L^{p}$ being a solution of the Yang-Baxter equation, 
\begin{equation}
L_{12}^{p}(w/z)L_{13}^{p}(w)R_{23}(z)=R_{23}(z)L_{13}^{p}(w)L_{12}^{p}(w/z)\
.  \label{RLL}
\end{equation}%
Obviously, this is sufficient to guarantee that $T$ and $Q$ commute.
Contrary, to Baxter \cite{BxBook} we do not at the moment require the
commutation relation $[Q^{p}(w),Q^{p}(z)]=0$. We will come back to this
point later.

The major achievement of Drinfel'd \cite{Drin} and Jimbo \cite{Jimbo} was to
show that the solutions of (\ref{RLL}) naturally arise from quantum groups.
Instead of solving the Yang-Baxter equation one solves the simpler
intertwining relation 
\begin{equation}
L^{p}(w)(\pi _{w}^{p}\otimes \pi _{1})\Delta (x)=\left[ \left( \pi
_{w}^{p}\otimes \pi _{1}\right) \Delta ^{\text{op}}(x)\right] L^{p}(w),\quad
x\in U_{q}(b_{+})\;.  \label{L}
\end{equation}%
Here $\pi _{w}^{p}$ denotes a suitably chosen representation. Its precise
form is for the moment not important, it will be specified explicitly below
(Section 4.2) when $q$ is a primitive root of unity. What matters at the
moment is that it depends on additional free parameters and that it is
similar to an evaluation representation. Both properties hold also true for
\textquotedblleft generic\textquotedblright\ $q$, the difference lies in the
fact that $\pi ^{p}$ is finite-dimensional when $q$ is a root of unity but
infinite-dimensional otherwise. In both cases the intertwiner $L^{p}$ has
been explicitly constructed, see \cite{RW02} and \cite{KQ}. However, when $%
q^{N}\neq 1$ then one has to specify how a meaningful definition of the
trace over an infinite-dimensional space in (\ref{Qp}) can be given. This
can be achieved by introducing quasi-periodic boundary conditions or for $%
|q|^{\pm 1}<1$ through the restriction to positive (negative) spin-sectors;
see \cite{KQ3} for details.

\bigskip
\begin{center}
\includegraphics[totalheight=2.5cm]{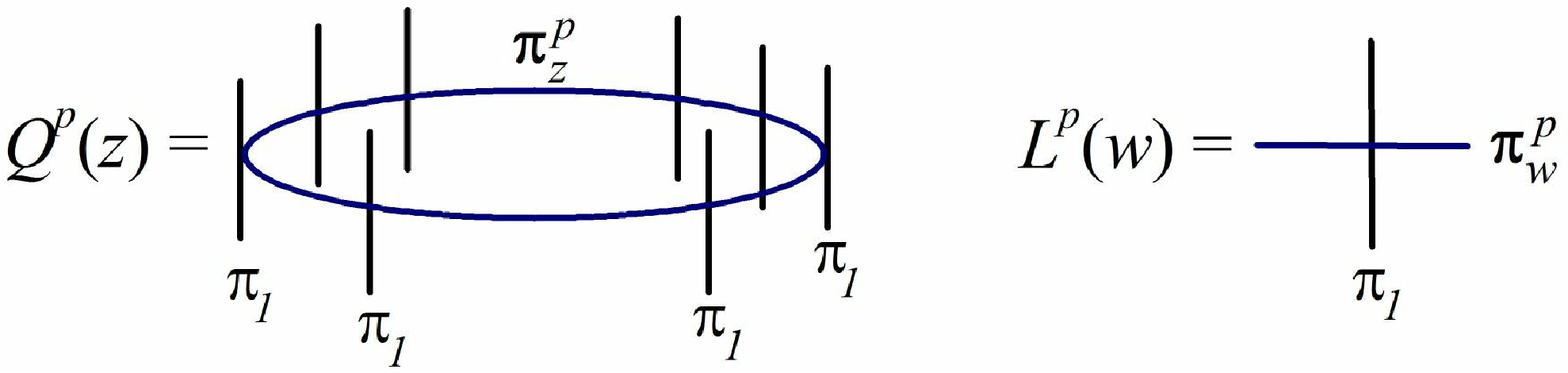}\\
{\small Figure 2. Graphical depiction of the auxiliary matrix and
the intertwiner\emph{\ L}.}\bigskip
\end{center}

\subsection{The derivation of the $TQ$ equation}

Having fixed our object $Q^{p}$ we now need to verify whether it actually
satisfies a functional equation with the transfer matrix which enables us to
express the eigenvalues of $T$ in terms of $Q^{p}$. Since by construction $T$
and $Q$ are built out of intertwiners the same applies to their operator
product, namely we have 
\begin{equation}
Q^{p}(w)T(z)=\limfunc{Tr}_{V_{0}\otimes V_{0^{\prime }}=\pi _{w}^{p}\otimes
\pi _{z}}L_{0M}^{p}(w)R_{0^{\prime }M}(z)\cdots L_{01}^{p}(w)R_{0^{\prime
}1}(z)  \label{QTprod}
\end{equation}%
with $L_{0m}^{p}(w)R_{0^{\prime }m}(z)$ being the intertwiner with respect
to the three-fold tensor product $[\pi _{w}^{p}\otimes \pi _{z}]\otimes \pi
_{1}$. In Figure 3 below the product of $Q^{p}$ and $T$ is diagrammatically
presented by the object to the utmost left. There are two horizontal lines
one representing the auxiliary space $V_{0}=\pi _{w}^{p}$ of the $Q$%
-operator and one $V_{0^{\prime }}=\pi _{z}$ the auxiliary space of the
transfer matrix. This interpretation now motivates to investigate the
properties of the tensor product representation $\pi _{w}^{p}\otimes \pi
_{z} $.

So far the parameter $w$ has been assumed to be free, we now fix it to a
value (depending on $z$ and the parameters $p$) such that the tensor product 
$\pi _{w}^{p}\otimes \pi _{z}$ becomes a reducible representation. That is,
it contains a proper subrepresentation of the quantum group which turns out
be similar to $\pi _{w}^{p}$ up to a possible shift in the parameters, $%
p\rightarrow p^{\prime }$ and $w\rightarrow w^{\prime }$. We explain
momentarily how the value $w$ and the subrepresentation $\pi _{w^{\prime
}}^{p^{\prime }}$ can be explicitly computed. Since we are dealing with a
non semi-simple algebra we can not expect that its reducible representations
decompose always into a direct sum of irreducible representations. Instead
we have to take the quotient space $\pi _{w}^{p}\otimes \pi _{z}/\pi
_{w^{\prime }}^{p^{\prime }}$ which once more turns out to be another
representation $\pi _{w^{\prime \prime }}^{p^{\prime \prime }}$. This
decomposition of the joint auxiliary space of $Q^{p}$ and $T$ is
conveniently summarized in the following non-split exact sequence, 
\begin{equation}
0\rightarrow \pi _{w^{\prime }}^{p^{\prime }}\overset{\imath }{%
\hookrightarrow }\pi _{w}^{p}\otimes \pi _{z}\overset{\tau }{\rightarrow }%
\pi _{w^{\prime \prime }}^{p^{\prime \prime }}\rightarrow 0\;.  \label{seq}
\end{equation}%
Here $\imath $ denote the inclusion map which identifies $\pi _{w^{\prime
}}^{p^{\prime }}$ and $\tau $ the projection map which sends $\pi
_{w^{\prime }}^{p^{\prime }}$ to zero and determines $\pi _{w^{\prime \prime
}}^{p^{\prime \prime }}$. Both maps need to be explicitly constructed. Once
this is done, the $TQ$ equation depicted in Figure 3 can be derived.

\bigskip
\begin{center}
\includegraphics[totalheight=2.75cm]{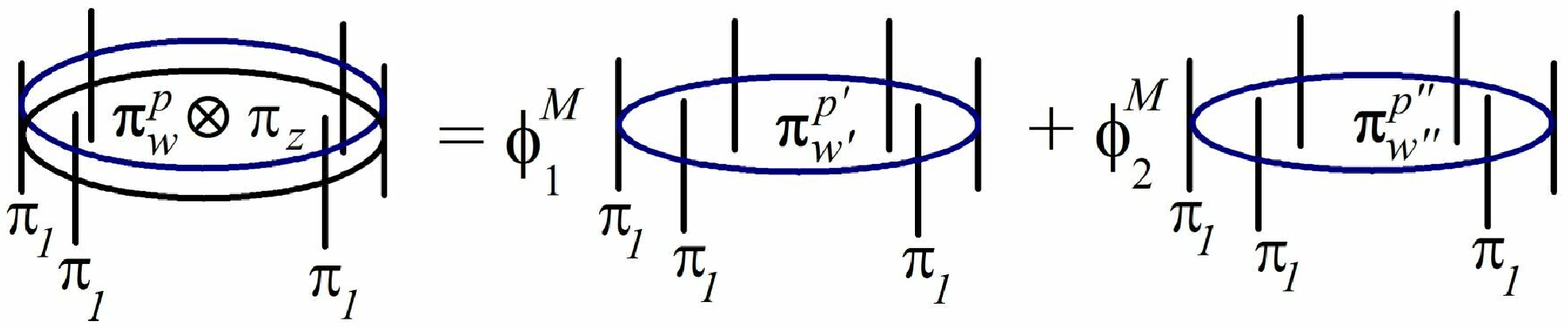}\\
{\small Figure 3. Graphical depiction of the \emph{TQ} equation.} \bigskip
\end{center}

Before we describe this derivation and the computation of the scalar
coefficients $\phi _{1,2}$ let us point out how one finds the value of $w$
and the subrepresentation $\pi _{w^{\prime }}^{p^{\prime }}$ which provide
the starting point. This information is extracted from the intertwiner of
the tensor product $\pi _{w}^{p}\otimes \pi _{z}$ which is simply $%
L^{p}(w/z) $, i.e. our building block of the auxiliary matrix. The tensor
product $\pi _{w}^{p}\otimes \pi _{z}$ is reducible if and only if $%
L^{p}(w/z)$ has a non-trivial kernel which coincides with the image of $\pi
_{w^{\prime }}^{p^{\prime }}$ under the inclusion map, i.e. we must have 
\begin{equation}
\det L^{p}(w/z)=0\text{\qquad and\qquad }\ker L^{p}(w/z)=\imath (\pi
_{w^{\prime }}^{p^{\prime }})
\end{equation}%
under an appropriate choice of $w$. This also determines the representation
space $\pi _{w^{\prime \prime }}^{p^{\prime \prime }}$ by taking the
quotient $\pi _{w}^{p}\otimes \pi _{z}/\pi _{w^{\prime }}^{p^{\prime }}$ as
pointed out before. In order to fully specify the inclusion and projection
map we now need to pick a suitable basis in the respective representation
spaces.

\subsubsection{The inclusion}

The inclusion $\imath :\pi _{w^{\prime }}^{p^{\prime }}\hookrightarrow \pi
_{w}^{p}\otimes \pi _{z}$ has so far only been characterized by identifying
its image. For the actual calculations and the derivation of the coefficient 
$\phi _{1}$ in the TQ equation one needs a concrete identification on the
level of basis vectors. This is particularly simple if $\pi _{w}^{p}\otimes
\pi _{z},\pi _{w^{\prime }}^{p^{\prime }}$ are highest or lowest weight
representations. For instance, let us assume that we have lowest weight
representations and denote the respective vectors by $\left\vert
0\right\rangle $ and $\left\vert 0\right\rangle ^{\prime }$ then we must
have $\pi _{w}^{p}(e_{1})\left\vert 0\right\rangle =\imath \lbrack \pi
_{w^{\prime }}^{p^{\prime }}(e_{1})\left\vert 0\right\rangle ^{\prime }]=0$.
The representation $\pi _{w^{\prime }}^{p^{\prime }}$ can then be generated
by the quantum group action via the identification 
\begin{equation}
\imath \lbrack \pi _{w^{\prime }}^{p^{\prime }}(x)\left\vert 0\right\rangle
^{\prime }]=(\pi _{w}^{p}\otimes \pi _{z})\Delta (x)\,\left\vert
0\right\rangle ,\quad \quad x\in U_{q}(\widehat{sl}_{2})\;.  \label{incD}
\end{equation}%
This relation is also used to determine the parameters $p^{\prime
},w^{\prime }$. As the quantum group action also fixes the intertwiner via (%
\ref{inter}) up to a possible normalization constant, the intertwiner of the
three-fold tensor product $\pi _{w}^{p}\otimes \pi _{z}\otimes \pi _{1}$
must coincide on the subspace $\imath (\pi _{w^{\prime }}^{p^{\prime
}})\otimes \pi _{1}$ with the intertwiner of $\pi _{w^{\prime }}^{p^{\prime
}}\otimes \pi _{1}$ up to a scalar multiple. That is, we have the identity 
\begin{equation}
L_{13}^{p}(w)R_{23}(z)(\imath \otimes 1)=\phi _{1}\,(\imath \otimes
1)L^{p^{\prime }}(w^{\prime }),  \label{LRi}
\end{equation}%
which is graphically depicted in Figure 4. The right-pointing fork
represents the inclusion map and each intersection on the various lines
stands for the respective intertwiners. (Note the similarity with the
bootstrap equation in factorizable scattering theories of QFT.)

\bigskip
\begin{center}
\includegraphics[totalheight=2.5cm]{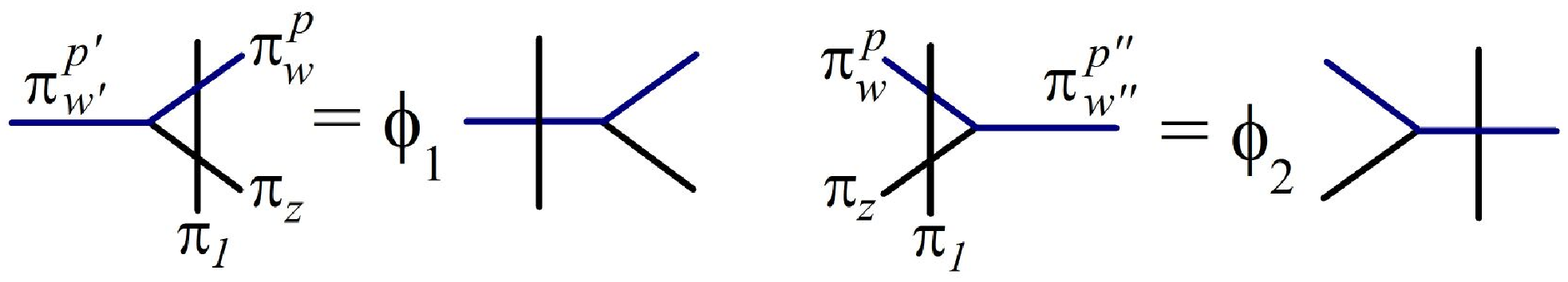}\\
{\small Figure 4. Graphical depiction of the equations (\protect\ref{LRi}) and (\protect\ref{pLR}).}\bigskip
\end{center}

The scalar multiple
occurring fixes the coefficient of the first term in the TQ equation and
must be explicitly calculated.

\subsubsection{The projection}

In the case of the projection map $\tau :\pi _{w}^{p}\otimes \pi
_{z}\rightarrow \pi _{w^{\prime \prime }}^{p^{\prime \prime }}$ we proceed
analogously. The only minor complication is that we have now to identify all
the vectors in the image of the inclusion map with the null vector in $\pi
_{w^{\prime \prime }}^{p^{\prime \prime }}$, i.e. the composition $\imath
\circ \tau $ vanishes. If $\tau $ is presented as a left-pointing fork then
its composition with $\imath $ yields the bubble diagram shown in Figure 5.%

\bigskip
\begin{center}
\includegraphics[totalheight=1cm]{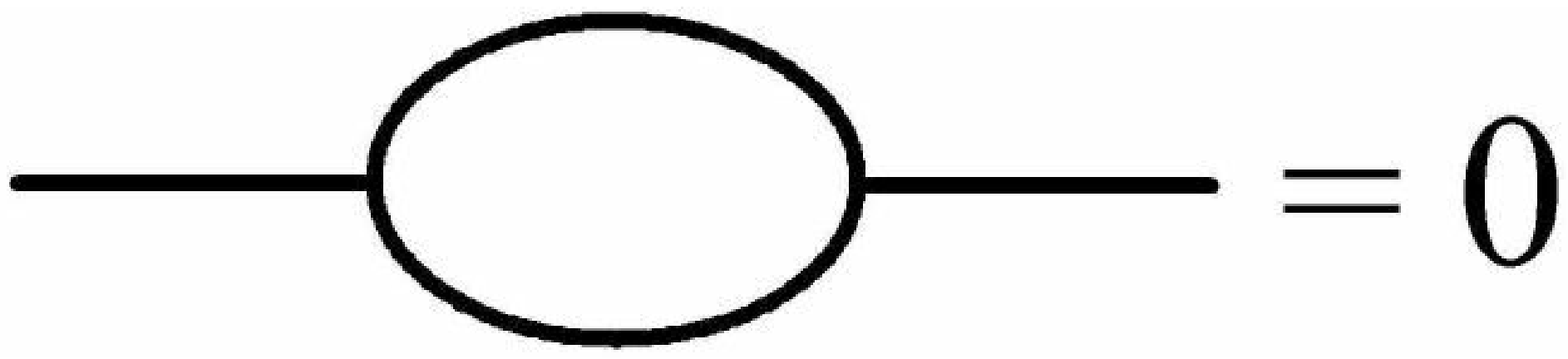}\\
{\small Figure 5. The
combination of inclusion and projection vanishes.}\bigskip
\end{center}

 Again, the construction
of the map $\tau $ is simplified when we are dealing with highest or lowest
weight representations and the counterpart of equation (\ref{incD}) is then 
\begin{equation}
\tau \lbrack (\pi _{w}^{p}\otimes \pi _{z})\Delta (x)\,\left\vert
0\right\rangle ]=\pi _{w^{\prime \prime }}^{p^{\prime \prime }}(x)\left\vert
0\right\rangle ^{\prime \prime },\quad \quad x\in U_{q}(\widehat{sl}_{2})\;.
\end{equation}%
By the same argument as before we must have the following identity for the
respective intertwiners 
\begin{equation}
(\tau \otimes 1)L_{13}^{p}(w)R_{23}(z)=\phi _{2}\,L^{p^{\prime \prime
}}(w^{\prime \prime })(\tau \otimes 1)\;.  \label{pLR}
\end{equation}%
The diagram corresponding to this equation is shown in Figure 4. Note that
the orientation matters as the coefficients $\phi _{1},\phi _{2}$ are in
general different.

Our findings can now be briefly summarized as follows. The intertwiner of
the three-fold tensor product $[\pi _{w}^{p}\otimes \pi _{z}]\otimes \pi
_{1} $ appearing under the trace in (\ref{QTprod}) can be written (in a
simplified notation) as the upper-triangular matrix 
\begin{equation*}
L_{13}^{p}R_{23}=\left( 
\begin{array}{cc}
\phi _{1}~L^{p^{\prime }} & \ast \\ 
0 & \phi _{2}~L^{p^{\prime \prime }}%
\end{array}%
\right)
\end{equation*}%
from which the $TQ$ equation (\ref{TQ}) is now easily deduced. To ensure
that the $TQ$ equation also holds on the level of eigenvalues we need to
show that all operators in (\ref{TQ}) simultaneously commute. This has been
done in \cite{KQ} for the root-of-unity case employing the intertwiners
constructed in \cite{BS90}. For $q^{N}\neq 1$ this has been indirectly shown
using the algebraic Bethe ansatz \cite{KQ3}. The stronger condition $%
[Q^{p}(z),Q^{p^{\prime }}(w)]=0$ which would correspond to Baxter's axiom
(v) in Chapter 9.5 of \cite{BxBook} is at the moment only rigorously proved
for some cases at roots of unity; see (\ref{Qcomm}) below.

\section{Roots of unity}

We now make contact with the special symmetries mentioned in the
introduction which are present when we set the deformation parameter $q$ of
the quantum group to be a primitive root of unity, $q^{N}=1$. We set $%
N^{\prime }=N/2$ when $N$ is even and $N^{\prime }=N$ when the order is odd.
Crucial for the following discussion is to appreciate that the full enriched
algebraic structure available can only be seen if one is directly at a root
of unity and not by simply taking the root-of-unity limit in the formulae
for \textquotedblleft generic\textquotedblright\ $q$. This is reflected on
the level of the underlying algebra by the fact that there exist now \emph{%
two} alternative, fundamentally different versions of the quantum group:

\begin{itemize}
\item[(1)] The unrestricted quantum group $U_{q}(\widehat{sl}_{2})$ \cite%
{dCKP}. This version has an enlarged centre compared to the case of
\textquotedblleft generic\textquotedblright\ $q$. The additional central
elements are generated by $\{e_{i}^{N^{\prime }},f_{i}^{N^{\prime
}},q^{N^{\prime }h_{i}}\}$ which can take non-zero values in the
representation theory at roots of unity. We will exploit this fact for the
construction of the auxiliary matrices.

\item[(2)] The restricted quantum group $U_{q}^{\text{res}}(\widehat{sl}%
_{2}) $ \cite{CP97}. The second version of the quantum group does not have
an enlarged centre and can be thought of as an algebra of derivations acting
on $U_{q}(\widehat{sl}_{2})$. For instance, one identifies the quantum group
generators at generic $q$ with the derivation 
\begin{equation*}
e_{i}^{(n)}\equiv e_{i}^{n}/[n]_{q}\rightarrow \lbrack e_{i}^{(n)},\cdot
~],\quad \lbrack n]_{q}=\frac{q^{n}-q^{-n}}{q-q^{-1}},\quad n\in \mathbb{N},
\end{equation*}%
which can be uniquely extended to the case $q^{N}=1$. In particular, this
still holds true when $n=0\func{mod}N^{\prime }$ and the subalgebra of the
restricted quantum group generated by these elements can be identified with
the non-deformed, classical affine algebra $U(\widehat{sl}_{2})$ via the
quantum Frobenius homomorphism, 
\begin{equation*}
F:U_{q}^{\text{res}}(\widehat{sl}_{2})\rightarrow U(\widehat{sl}_{2}),\quad
e_{i}^{(n)}\rightarrow \left\{ 
\begin{array}{cc}
\bar{e}_{i}^{n/N^{\prime }}, & n=0\func{mod}N^{\prime } \\ 
0, & \text{otherwise}%
\end{array}%
\right. \;.
\end{equation*}%
This is the symmetry algebra of the six-vertex model at roots of unity
discovered by Deguchi, Fabricius and McCoy in \cite{DFM}.
\end{itemize}

\noindent We now discuss the various aspects of these two versions and their
relation to the six-vertex model at roots of unity separately.

\subsection{The affine symmetry algebra: the restricted quantum group}

Let us briefly review the affine symmetry algebra of the six-vertex transfer
matrix at roots of unity discovered in \cite{DFM}. Therein the symmetry was
established based on the Temperley-Lieb algebra, however,\ we recall here
the simplified proof given in \cite{Ktw} based on the intertwining property
of the six-vertex monodromy matrix 
\begin{equation}
\mathbf{T}=R_{0M}\cdots R_{01}=\left( 
\begin{array}{cc}
A & B \\ 
C & D%
\end{array}%
\right) \;.  \label{mom}
\end{equation}%
Namely, because of our earlier choice of the coproduct (\ref{cop}) we now
have 
\begin{equation}
\mathbf{T\,}\left( \pi _{z}\otimes \pi _{\mathcal{H}}\right) \Delta
(x)=[\left( \pi _{z}\otimes \pi _{\mathcal{H}}\right) \Delta ^{\text{op}%
}(x)]\,\mathbf{T}\quad \text{with}\quad \pi _{\mathcal{H}}=\pi _{1}^{\otimes
M},\quad x\in U_{q}(\widehat{sl}_{2})\;.  \label{mominter}
\end{equation}%
From this relation it is straightforward to derive the commutation relations
between the quantum group generators and the matrix elements of $\mathbf{T}$%
. For instance, we find for the transfer matrix $T=A+D$ and $x=e_{1}$, 
\begin{equation}
\pi _{\mathcal{H}}(e_{1}^{n})T=(q^{n}A+q^{-n}D)\,\pi _{\mathcal{H}%
}(e_{1}^{n})+[n]_{q}(1-q^{2S^{z}})C\,\pi _{\mathcal{H}}(e_{1}^{n-1})\;.
\label{Tb+}
\end{equation}%
Analogous relations hold for the remaining quantum group generators; see 
\cite{Ktw}. As (\ref{mominter}), (\ref{Tb+}) hold true for \textquotedblleft
generic\textquotedblright\ $q$ we infer for roots of unity $q^{N}=1$ and the
restricted quantum group generators with $n=N^{\prime }$ that 
\begin{equation}
\pi _{\mathcal{H}}(e_{1}^{(N^{\prime })})T=q^{N^{\prime }}T\,\pi _{\mathcal{H%
}}(e_{1}^{(N^{\prime })})\quad \text{iff\quad }2S^{z}=0\func{mod}N\;.
\label{loop}
\end{equation}%
The above condition defines the commensurate spin-sectors of the affine
symmetry algebra, only there the second term on the right hand side of (\ref%
{Tb+}) vanishes. This does not mean that there are no degeneracies outside
these spin-sectors. In fact, it has been shown \cite{DFM} for the free
fermion case, $q^{4}=1$, and by a numerical construction for $N=3$ that
certain projection operators have to be introduced to obtain the symmetry
algebra in all sectors. The general case is still open. Notice, however,
that for lattices with $M$ odd and even roots of unity there are no
commensurate sectors \ and, indeed there are no extra degeneracies in the
spectrum of the transfer matrix besides spin-reversal. We will come back to
this point below.

In \cite{Ktw} it has been established that for a special choice of
quasi-periodic boundary condition the affine symmetry (\ref{loop}) extends
to all spin-sectors without the need of projection operators, although it is
broken down to $U_{q}(b_{\pm })$ outside the commensurate sectors.

\subsection{Auxiliary matrices: the non-restricted quantum group}

The construction of the auxiliary matrices $Q^{p}$ relies on an appropriate
choice of the representation $\pi _{w}^{p}$. We now specify this
representation as an irreducible representation of the non-restricted
quantum group which has an enlarged centre. We make the (slightly
restrictive) assumption that $\pi _{w}^{p}$ is an evaluation representation,
i.e. similarly as in the case of the transfer matrix we set $\pi
_{w}^{p}=\pi ^{p}\circ \,$ev$_{w}$ with ev being Jimbo's evaluation
homomorphism (\ref{ev}) and $\pi ^{p}$ a representation of the finite
quantum group $U_{q}(sl_{2})$. This allows us to make contact with the
discussion in \cite{dCKP}.

The finite-dimensional irreducible representations of $U_{q}(sl_{2})$ at a
primitive root of unity $q^{N}=1$ are determined (up to isomorphism) by the
values of the following central elements, 
\begin{equation}
\mathbf{x}=\left[ (q-q^{-1})e\right] ^{N^{\prime }},\quad \mathbf{y}=\left[
(q-q^{-1})f\right] ^{N^{\prime }},\quad \mathbf{z}=(q^{h})^{N^{\prime }}
\label{Z0}
\end{equation}%
and the Casimir operator $\mathbf{c}=q^{h+1}+q^{-h-1}+(q-q^{-1})^{2}ef~.$
Each representation now assigns to these central elements four complex
numbers which comprise our parameters $p$, i.e. $(\mathbf{x},\mathbf{y},%
\mathbf{z},\mathbf{c})\rightarrow p=(p_{1},p_{2},p_{3},p_{4})\in \mathbb{C}%
^{4}$. An example for such a representation $\pi ^{p}$ over the vector space 
$\mathbb{C}^{N^{\prime }}$ is given by \cite{RA,dCKP}, 
\begin{eqnarray}
\pi ^{p}(q^{h})\left\vert n\right\rangle &=&\lambda q^{-2n}\left\vert
n\right\rangle ,\quad \pi ^{p}(f)\left\vert n\right\rangle =\left\vert
n+1\right\rangle ,\quad \pi ^{p}(f)\left\vert N^{\prime }-1\right\rangle
=\zeta \left\vert 0\right\rangle ,  \notag \\
\pi ^{p}(e)\left\vert n\right\rangle &=&\xi _{n}\left\vert n-1\right\rangle
,\quad \pi ^{p}(e)\left\vert 0\right\rangle =\xi \left\vert N^{\prime
}-1\right\rangle  \label{pip}
\end{eqnarray}%
where $n=0,1,2,...,N^{\prime }-1$ and 
\begin{equation}
\xi _{n}:=\frac{\lambda q+\lambda ^{-1}q^{-1}-\lambda q^{-2n+1}-\lambda
^{-1}q^{2n-1}}{(q-q^{-1})^{2}}+\xi \zeta ,\quad n>0\;.
\end{equation}%
The values $p=p(\xi ,\zeta ,\lambda )$ of the central elements (\ref{Z0})
are given by 
\begin{equation}
\pi ^{p}(\mathbf{x})=(q-q^{-1})^{N^{\prime }}\xi \prod_{n=1}^{N^{\prime
}-1}\xi _{n},\quad \pi ^{p}(\mathbf{y})=\zeta (q-q^{-1})^{N^{\prime }},\;\pi
^{p}(\mathbf{z})=\lambda ^{N^{\prime }}  \notag
\end{equation}%
and for the Casimir element one finds $\pi ^{p}(\mathbf{c})=q\lambda
+q^{-1}\lambda ^{-1}+(q-q^{-1})^{2}\xi \zeta \;.$ Note that the
representation (\ref{pip}) is cyclic when $\xi ,\zeta \neq 0$, i.e. there
are no highest or lowest weight vectors.

Having specified the representation $\pi ^{p},$ and therefore also the
evaluation representation $\pi _{w}^{p}$ of the affine quantum group, one
can construct the intertwiner $L^{p}$ via (\ref{L}), which forms the
building block of the auxiliary matrix (\ref{Qp}). However, due to the
enlarged centre of the quantum group this intertwiner might only exist for
special values of the central elements, in other words the concept of the
universal R-matrix in (\ref{inter}) breaks down; see the discussion in \cite%
{KQ} and references therein. For instance, we find that for even roots of
unity we must set $\xi =\zeta =0$ in order to find a solution of (\ref{L}).

For odd roots of unity, however, there are no restrictions on $\xi ,\zeta $
or $\lambda $ and for $\xi ,\zeta \neq 0$ the resulting $Q$-operators do not
preserve the spin. We comment on this case in more detail, as it provides us
with an interesting geometric picture of the solutions to the $TQ$ equation (%
\ref{TQ}).

\subsubsection{A geometric picture for $N$ odd}

As the reader might have already noticed from the representation (\ref{pip})
the central elements (\ref{Z0}) and the Casimir operator are not
algebraically independent. In fact, their values ought to lie on a
three-dimensional hypersurface Spec\thinspace $Z$ specified by the following
identity, 
\begin{equation}
\text{Spec}\,Z:\quad \mathbf{xy}+\mathbf{z}+\mathbf{z}^{-1}=\prod\limits_{%
\ell =0}^{N-1}\left( \mathbf{c}+q^{\ell }+q^{-\ell }\right) -2\;.
\label{cxyz}
\end{equation}%
Here we now interpret $(\mathbf{x},\mathbf{y},\mathbf{z},\mathbf{c})$ as $%
\mathbb{C}$-numbers which have to solve (\ref{cxyz}). To each solution, i.e.
a point on the hypersurface, we can then associate a representation $\pi
^{p} $ respectively $\pi _{w}^{p}$ giving rise to a solution $Q^{p}$ of the
operator functional equation (\ref{TQ}). So far we have not specified the
points $p^{\prime },p^{\prime \prime }$ appearing in the $TQ$ equation. To
this end recall from \cite{dCKP} that Spec$\,Z$ has locally the structure of
an $N$-fold covering space over the base manifold Spec$~Z_{0}=\{\mathbf{x},%
\mathbf{y},\mathbf{z}\}=\mathbb{C}^{3}$. The three points appearing in (\ref%
{TQ}) lie in the same fibre and decomposing $\mathbf{c}=\mu +\mu ^{-1}$ are
explicitly given by 
\begin{equation}
p=(\mathbf{x},\mathbf{y},\mathbf{z},\mu +\mu ^{-1}),\quad p^{\prime }=(%
\mathbf{x},\mathbf{y},\mathbf{z},\mu q+\mu ^{-1}q^{-1}),\quad p^{\prime
\prime }=(\mathbf{x},\mathbf{y},\mathbf{z},\mu q^{-1}+\mu ^{-1}q)\;.
\label{ppp}
\end{equation}%
See Figure 6 for a simplified graphical depiction. The values for the
evaluation parameters and coefficients $\phi _{1,2}$ are \cite{KQ} 
\begin{equation}
w=z/\mu ,\quad w^{\prime }=zq/\mu ,\quad w^{\prime \prime }=zq^{-1}/\mu
,\quad \phi _{1}(z)=\phi _{2}(zq^{-2})=z-1\;.  \label{www}
\end{equation}%
The motivation for making this connection lies in the rich structure of the
hypersurface (\ref{cxyz}) described in \cite{dCKP}.

\bigskip
\begin{center}
\includegraphics[totalheight=8cm]{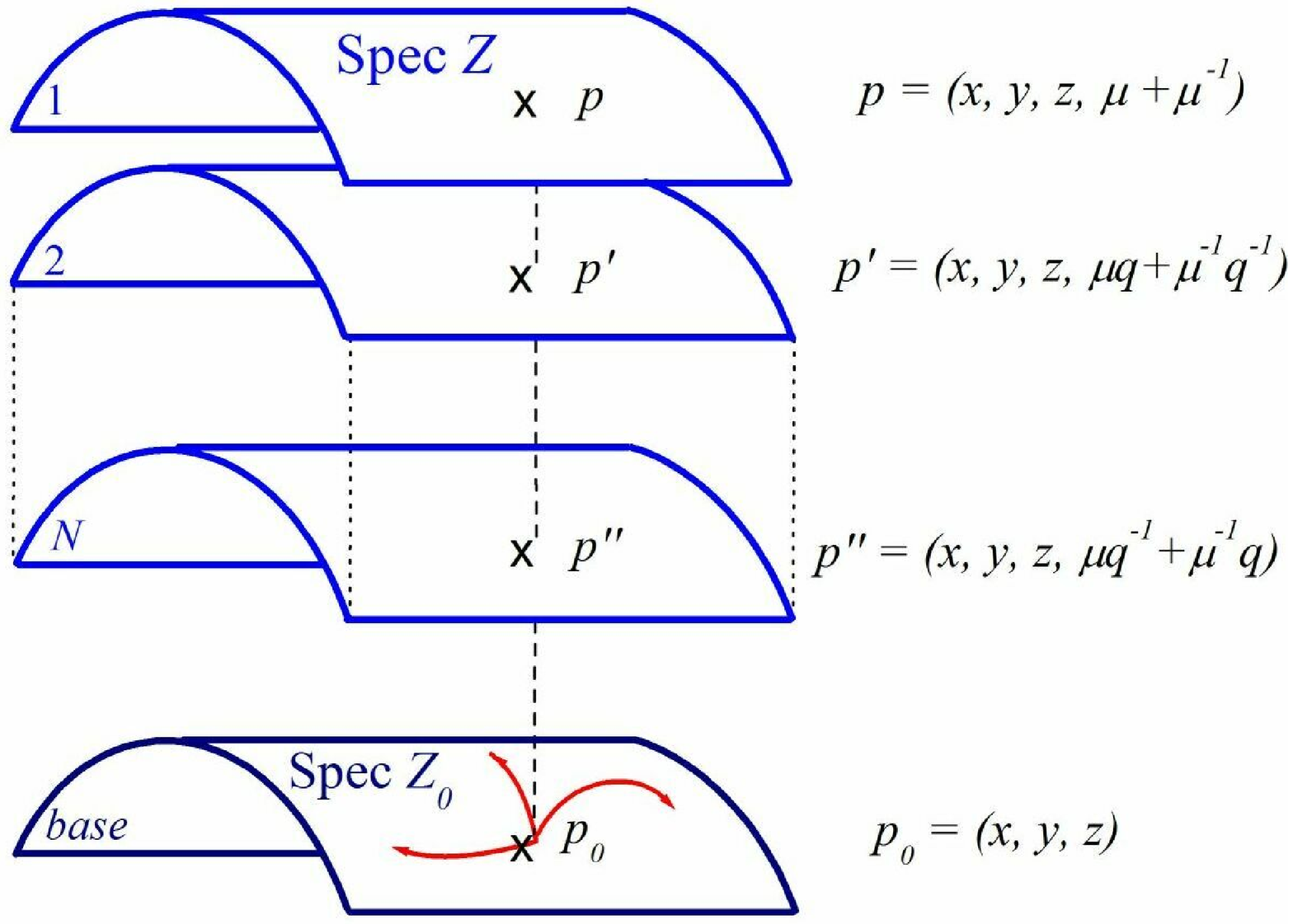}\\
{\small Figure 6. A simplified picture of the hypersurface (%
\protect\ref{cxyz}).}\bigskip
\end{center}

Spec\thinspace $Z$ is
endowed with an infinite-dimensional group action $G$, called the quantum
coadjoint action, which induces holomorphic transformations in the
coordinates $(\mathbf{x},\mathbf{y},\mathbf{z},\mathbf{c})$ and acts
transitively on the hypersurface. This action can be carried over to the
auxiliary matrices (\ref{Qp}). However, what is missing at the moment is an
implementation on the lattice, i.e. a map $D$ from the group $G$ of
transformations into the matrices acting on the quantum space $\pi _{%
\mathcal{H}}$ such that 
\begin{equation}
D:G\rightarrow \limfunc{End}\pi _{\mathcal{H}},\quad
D(g)Q^{p}D(g^{-1})=Q^{g\cdot p},  \label{D}
\end{equation}%
where $g\cdot p$ is the point on the hypersurface obtained under the
coordinate transformation $g\in G$. The map (\ref{D}) might provide the key for the construction of
the symmetry algebra (\ref{loop}) outside the commensurate sectors. It would
also simplify the calculation of the spectrum of the general set of
auxiliary matrices $Q^{p}$. As of yet the spectrum has only been calculated
for representations with $\mathbf{x}=\mathbf{y}=0$, which are called
nilpotent and which we discuss next.

\subsection{The spectrum for nilpotent representations}

We now lift the temporary restriction to odd roots of unity but, henceforth,
shall only consider representations $\pi ^{\mu }\equiv \pi ^{p_{\mu }}$ in
the set $\{p_{\mu }=(\mathbf{x}=0,\mathbf{y}=0,\mathbf{z}=\mu ^{-N^{\prime
}},\mu +\mu ^{-1}):\mu \in \mathbb{C}\}$. Note that the relations (\ref{ppp}%
) and (\ref{www}) remain true in the limit $\mathbf{x},\mathbf{y}\rightarrow
0$. Denoting the associated auxiliary matrices by $Q_{\mu }\equiv Q^{p_{\mu
}}$ it has been proved in \cite{KQ2} for $N=3$ and in \cite{KQ5} for $N=4,6$
that the following commutation relation holds, 
\begin{equation}
\lbrack Q_{\mu }(z),Q_{\nu }(w)]=0,\quad \forall z,w,\mu ,\nu \in \mathbb{C}%
\quad .  \label{Qcomm}
\end{equation}%
The proofs given rely on the explicit construction of the quantum group
intertwiners with respect to the tensor product $\pi _{w}^{\mu }\otimes \pi
_{1}^{\nu }$. Although the construction for the general case is an open
problem the necessary condition for the intertwiner to exist are satisfied
and numerical computations for $N=5,7,8$ up to $M=11$ confirm that it is
correct also for higher roots of unity. From (\ref{Qcomm}) one can deduce
two cardinal facts \cite{KQ2,KQ4}:

\begin{enumerate}
\item The auxiliary matrix $Q_{\mu }$ is normal and hence diagonalizable.

\item The eigenvalues of $Q_{\mu }(z)$ are polynomials in $z$ which are at
most of degree $M$.
\end{enumerate}

\noindent The additional information required to determine the
characteristics of the eigenvalues is yet again derived from another
operator functional equation \cite{KQ4},%
\begin{equation}
Q_{\mu }(z\mu ^{2}q^{2})Q_{\nu }(z)=(zq^{2}-1)^{M}Q_{\mu \nu q}(z\mu
^{2}q^{2})+Q_{\mu \nu q^{N^{\prime }+1}}(z\mu ^{2}q^{2})T^{(N^{\prime
}-1)}(zq^{2}),  \label{QQQ}
\end{equation}%
which similar as in the case of the $TQ$ equation is deduced from the
decomposition of a tensor product \cite{KQ4},

\begin{equation}
0\rightarrow \pi _{w^{\prime }}^{\mu ^{\prime }}\overset{\imath }{%
\hookrightarrow }\pi _{w}^{\mu }\otimes \pi _{1}^{\nu }\overset{\tau }{%
\rightarrow }\pi _{w^{\prime \prime }}^{\mu ^{\prime \prime }}\otimes \pi
_{z^{\prime }}^{(N^{\prime }-2)}\rightarrow 0\ .
\end{equation}%
Here $T^{(N^{\prime }-1)}$ denotes the fusion matrix of degree $N^{\prime
}-1 $, i.e. the analogue of the transfer matrix for the six-vertex model
with spin $(N^{\prime }-2)/2$. The various parameters appearing in the
representations are not all independent but satisfy the relations 
\begin{equation}
w=\mu \nu q^{2},\quad \mu ^{\prime }=\mu \nu q,\quad w^{\prime }=\mu q,\quad
\mu ^{\prime \prime }=\mu \nu q^{-N^{\prime }+1},\quad w^{\prime \prime
}=\mu q^{N^{\prime }+1},\quad z^{\prime }=\nu q^{N^{\prime }+1}\;.
\label{var}
\end{equation}%
We do not want to go into the details of the derivation as it follows the
analogous steps as detailed in Section 3. However, it provides a more
complicated example and underlines the general nature of the approach which
not only applies to the $TQ$ equation. See Figure 7 for a graphical
depiction of the equation.

\bigskip
\begin{center}
\includegraphics[totalheight=2.75cm]{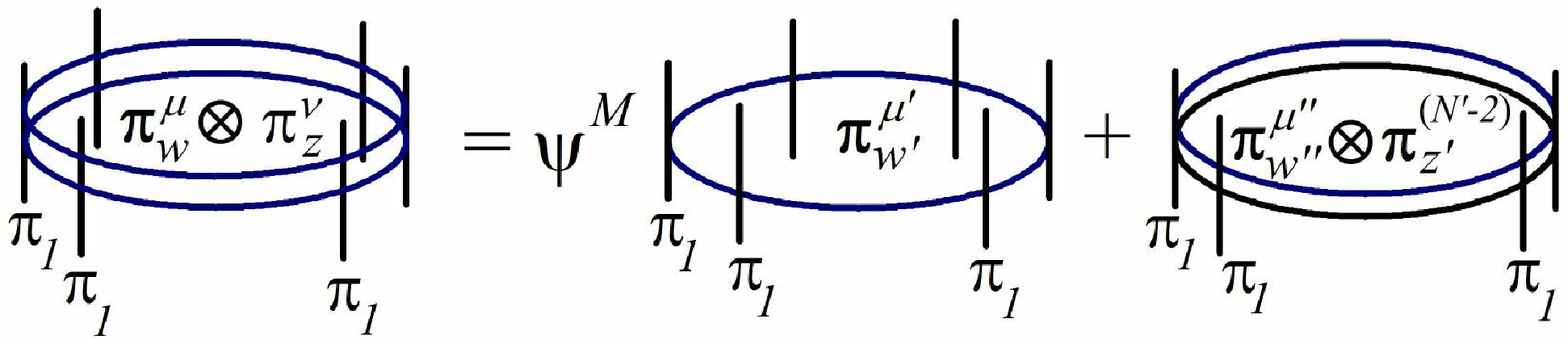}\\
{\small Figure 7. Graphical depiction of the functional equation (%
\protect\ref{QQQ}).}\bigskip
\end{center}

The result on the structure of the eigenvalues, which we denote by the same
symbol as the corresponding operator, can be summarized as follows. It
factorizes into two polynomials,%
\begin{equation}
Q_{\mu }(z)=Q^{+}(z)Q_{\mu }^{-}(z),  \label{factor}
\end{equation}%
one of which, $Q^{+},$ does not depend on the free parameter $\mu $. The
other factor $Q_{\mu }^{-}$ can be expressed through the first one in the
special limit $\mu \rightarrow q^{-N^{\prime }},$%
\begin{equation}
Q^{-}(z):=\lim_{\mu \rightarrow q^{-N^{\prime }}}Q_{\mu
}^{-}(z)=\,q^{(N^{\prime }+1)s}Q^{+}(z)\sum_{\ell =1}^{N^{\prime }}\frac{%
q^{-2\ell s}(zq^{2\ell }-1)^{M}}{Q^{+}(zq^{2\ell })Q^{+}(zq^{2\ell -2})}\ .
\label{inv}
\end{equation}%
It is not immediately apparent that the right hand side of this equation
defines a polynomial. This follows from the $TQ$ equation (\ref{TQ}) which
implies that the zeroes of $Q^{+}$ satisfy the six-vertex Bethe ansatz
equations. The parameter $s$ in (\ref{inv}) can be identified with 
\begin{equation}
s=2n_{0}+S^{z}\func{mod}N^{\prime }\ ,
\end{equation}%
where $n_{0}$ denotes the number of Bethe roots which vanish in the root of
unity limit $q^{N}\rightarrow 1$. That is, in general we have less Bethe
roots than in the case when $q^{N}\neq 1$, $\deg Q^{+}\leq M/2-S^{z}$
instead of $\deg Q^{+}=M/2-S^{z}$

\subsubsection{Case-by-case discussion}

In order to further characterize the spectrum and explain how the free
parameter $\mu $ enters we have now to distinguish various cases. As we
already saw earlier the loop symmetry (\ref{loop}) is absent for lattices
with an number of columns and even roots of unity. This will be reflected in
the spectrum of the auxiliary matrices. It needs to be emphasized that the
results of \cite{KQ4} presented here are in accordance with the findings of
Fabricius and McCoy in the eight-vertex case, see \cite{FM8v3} and \cite%
{FMproc}. We will comment further on this in the conclusion.\medskip

\noindent $M$\textbf{\ even, }$N$\textbf{\ arbitrary.} In this case the
second factor takes the following form%
\begin{equation}
Q_{\mu }^{-}(z)=\mathcal{N}_{\mu }\text{\thinspace }z^{n_{\infty
}}Q^{+}(z\mu ^{-2})\,P_{S}(z^{N^{\prime }}),\quad P_{S}(z^{N^{\prime
}})=\tprod_{i=1}^{n_{S}}(1-z^{N^{\prime }}a_{i})\ .
\end{equation}%
Here $\mathcal{N}_{\mu }$ is a normalization constant which only depends on $%
\mu $ and $q$, the power of the polynomial depends on the number of Bethe
roots which have either vanished or gone off to infinity in the root of
unity limit. The last factor $P_{S},$ which drops out of the $TQ$ equation (%
\ref{TQ}), determines the degeneracy of the corresponding eigenvalue of the
transfer matrix $T$. Denote by $V_{T}$ the associate degenerate eigenspace
of the transfer matrix then we have%
\begin{equation}
\dim V_{T}=2^{\deg P_{S}}\ .  \label{mult}
\end{equation}%
This result follows from the fact that the auxiliary matrix $Q_{\mu }$ lifts
the degeneracy of the transfer matrix. Inside the degenerate eigenspace $%
V_{T}$ the set of eigenvalues of $Q_{\mu }$ varies only through the
dependence of each zero $a_{i}$ of $P_{S}$ on the parameter $\mu $. There
are only two choices: either $a_{i}$ depends on $\mu $ through a simple
multiplicative factor $\mu ^{2N^{\prime }}$ or it does not depend on it at
all. Hence, the maximal number of eigenvectors of the auxiliary matrix $%
Q_{\mu }$ in a degenerate eigenspace $V_{T}$ of the transfer matrix is given
by (\ref{mult}).

In addition, it has been shown for several examples with $N=3$ in \cite{KQ2}
that the zeroes $\{a_{i}\}$ coincide with the evaluation parameters of the
loop algebra, i.e. $P_{S}$ has been identified with the Drinfel'd
polynomial. To prove this assertion for general $N$ a deeper understanding
of the highest weight vectors of the symmetry algebra (\ref{loop}) is
desirable; see \cite{DFM, FM01a, FM01b}.\medskip

\noindent $M$\textbf{\ odd, }$N$\textbf{\ odd.} For odd chains the just
presented picture remains true with the possible exception that some
eigenvalues of the auxiliary matrix may now vanish, i.e. we have the two
possible cases%
\begin{equation*}
\mathcal{N}=0\quad \text{or\quad }Q_{\mu }^{-}(z)=\mathcal{N}_{\mu }\text{%
\thinspace }z^{n_{\infty }}Q^{+}(z\mu ^{-2})\,P_{S}(z^{N^{\prime }})\text{%
\quad with\quad }\mathcal{N}\neq 0\;.
\end{equation*}%
The vanishing of some eigenvalues seems at first to be a serious drawback.
However, the vanishing of the eigenvectors occurs only for singlet states,
i.e. non-degenerate eigenvectors of the transfer matrix. The number of such
vectors rapidly decreases as $M$ starts to exceed the order $N$ \cite{KQ4}.
More importantly, since the relation (\ref{inv}) remains true and $Q^{+}$ is
non-vanishing one can derive a set of difference equations which yield
constraints on the Bethe roots, i.e. the zeroes of $Q^{+}$ \cite{KQ4}. These
constraints are polynomial equations of order $N-2$ in contrast to the Bethe
ansatz equation which are of order $M$. For $N=3$ these can be explicitly
solved and one obtains Stroganov's result \cite{S01}; see \cite{KQ4} and
references therein for further details.\medskip

\noindent $M$\textbf{\ odd, }$N$\textbf{\ even.} For the last case we
discuss, the spectrum of the six-vertex transfer matrix shows no
degeneracies except for spin-reversal symmetry. Now $Q_{\mu }^{-}$ does not
factorize as in the previous cases and we have $Q_{\mu }^{-}(z)=Q^{-}(z\mu
^{-2})$ up to some normalization factor which is not important. The second
factor $Q^{-}$ now constitutes a second linearly independent solution to
Baxter's $TQ$ equation on the level of eigenvalues and its zeroes are the
solutions to the Bethe ansatz equations below the equator. This is an
explicit construction of the analogous scenario discussed in \cite{PrSt}
away from roots of unity. The two linearly independent solutions are bound
to satisfy a Wronskian equation,

\begin{equation}
q^{S^{z}}Q^{+}(zq^{2})Q^{-}(z)-q^{-S^{z}}Q^{+}(z)Q^{-}(z)=\text{const.\ }%
(zq^{2}-1)^{M},  \label{W}
\end{equation}%
where the degree of the polynomials now obey 
\begin{equation*}
\deg Q^{\pm }=\frac{M}{2}\mp S^{z}\ .
\end{equation*}%
Note that the Wronskian equation (\ref{W}) implies the six-vertex Bethe
ansatz equations; see \cite{KQ4,KQ5} for details.\medskip

While we have discussed here the factorization of the auxiliary matrix $%
Q_{\mu }$ into the factors $Q^{\pm }$ only on the level of eigenvalues,
there is a simplified construction which assigns to each of the factors a
proper operator; see \cite{KQ5} for details.

\section{Concluding Remarks}

We would like to stress once more that the representation theoretic
construction of $Q$-operators and the derivation of the $TQ$ equation
presented in Section 3 apply to the case of \textquotedblleft
generic\textquotedblright\ $q$ as well as roots of unity. Although in this
overview the representation $\pi ^{p}$ in Definition (\ref{Qp}) has only
been specified for $q^{N}=1$ the case $q^{N}\neq 1$ has been discussed in 
\cite{RW02}; see also \cite{KQ3} for a discussion resolving the convergence
problems with an infinite-dimensional representation $\pi ^{p}$. At the
moment we are still missing the analogue of (\ref{Qcomm}) and (\ref{QQQ})
when $q^{N}\neq 1$. The method can also be extended to more complicated
models than the six-vertex one. For instance, those associated with higher
rank algebras. However, one has then to account for the possibility of a
more complicated decomposition of the tensor products, similar to the one we
encountered in the derivation of (\ref{QQQ}).

The other obvious target for a generalization of this method is the
eight-vertex model. The investigation \cite{FM8v,FM8v2,FM8v3} of Fabricius
and McCoy has already extended in great detail our knowledge of the spectrum
of the eight-vertex model and Baxter's 1972 solution of the $TQ$ equation.
Their findings match closely the six-vertex picture summarized in this
article; see also \cite{FMproc}. However, because of the intricate algebraic
form of Baxter's 1972 eight-vertex $Q$-operator \cite{Bx72} it is difficult
to take directly the trigonometric limit and obtain a well-defined
six-vertex $Q$-operator. At the moment we can therefore match the eight and
six-vertex results \emph{on the level of eigenvalues} only. Then the factor $%
Q^{+}$ in (\ref{factor}) should be identified as the six-vertex analogue of
Baxter's $Q$-operator in \cite{Bx72}. The other factor, $Q^{-}$, corresponds
to the same eight-vertex $Q$ but when it is evaluated in a different regime.
Notice that through a recent refined construction presented in \cite{KQ5}
the factorization (\ref{factor}) can also be made on the level of operators
for the six-vertex model. What we are missing at the moment is a feasible
elliptic construction of the $Q$-operator which allows one to carry out the
trigonometric limit directly in the definition of the operator and to derive
the $TQ$ equation as well as the eight-vertex analogue of the essential
identity (\ref{inv}) discussed in \cite{FM8v,FM8v2} in a similar manner as
presented here. \medskip

\noindent \textsc{Acknowledgement}. The author would like to thank M.
Shiraishi and T. Miwa for the opportunity to present his results at the RIMS
workshop "Progress in Exactly Solvable Lattice Models", Kyoto, July 2004 and
M. Jimbo and B.M. McCoy for numerous insightful discussions. This work has
been financially supported by the EPSRC Grant GR/R93773/01 and a University
Research Fellowship of the Royal Society.


\begin{thebibliography}{99}
\bibitem{DFM} {\small T. Deguchi, K. Fabricius and B. M. McCoy, \emph{J.
Stat. Phys}. \textbf{102} (2004) 701}

\bibitem{FM01a} {\small K. Fabricius and B. M. McCoy, \emph{J. Stat. Phys}. 
\textbf{103} (2001) 647}

\bibitem{FM01b} {\small K. Fabricius and B. M. McCoy MathPhys Odyssey 2001
(Progress in Mathematical Physics 23) ed M. Kashiwara and T. Miwa (Boston:
Birkhauser) pp 119}

\bibitem{FM8v} {\small K. Fabricius and B. M. McCoy, \emph{J. Stat. Phys.} 
\textbf{111} (2003) 323}

\bibitem{FM8v2} {\small K. Fabricius and B. M. McCoy, \emph{Publ. RIMS} 
\textbf{40} (2004) 905}

\bibitem{FM8v3} {\small K. Fabricius and B. M. McCoy, \emph{New Developments
in the Eight Vertex Model II. Chains of odd length.} cond-mat/0410113}

\bibitem{FMproc} {\small K. Fabricius and B. M. McCoy, \emph{Root of unity
symmetries in the 8 and 6 vertex models.} cond-mat/0411419}

\bibitem{Bx72} {\small R. J. Baxter, \emph{Ann. Phys., NY} \textbf{70}
(1972) 193}

\bibitem{Bx73} {\small R. J. Baxter, \emph{Ann. Phys., NY} \textbf{76}
(1973) 1--24; 25--47; 48--71}

\bibitem{BxBook} {\small R. J. Baxter Exactly Solved Models in Statistical
Mechanics (London: Academic Press) 1982}

\bibitem{Lieb67} {\small E.H. Lieb \emph{Phys. Rev}. \textbf{162} (1967)
162; \emph{Phys. Rev. Lett}. \textbf{18} (1967) 1046;\emph{\ Phys. Rev. Lett}%
. \textbf{19} (1967) 108}

\bibitem{St67} {\small B. Sutherland, \emph{Phys. Rev. Lett.} \textbf{19}
(1967) 103}

\bibitem{FSZ} {\small V.~Fridkin, Y. Stroganov, D. Zagier, \emph{J. Phys. A:
Math. Gen}. \textbf{33} (2000) L125}

\bibitem{S01} {\small Y. Stroganov, \emph{J. Phys. A: Math. Gen}.\textbf{\ 34%
} (2001) L179}

\bibitem{BM} {\small V.V. Bazhanov, V.V. Mangazeev, Eight-vertex model and
non-stationary Lame equation, hep-th/0411094}

\bibitem{KWLZ} {\small I. Krichever, O. Lipan, P. Wiegmann, A. Zabrodin, 
\emph{Comm. Math. Phys.} \textbf{188} (1997) 267}

\bibitem{PrSt} {\small G. P. Pronko and Y. Stroganov, \emph{J. Phys. A:
Math. Gen.} \textbf{32} (1999) 2333}

\bibitem{Ku96} {\small G. Kuperberg, \emph{Internat. Math. Res. Notices} 
\textbf{3} (1996) 139}

\bibitem{RS} {\small A.V. Razumov and Yu. G. Stroganov, \emph{J. Phys.
A:~Math. Gen.} \textbf{34} (2001) 3185; On refined enumerations of some
symmetry classes of ASMs, math-ph/0312071.}

\bibitem{BGN} {\small M. T. Batchelor, J. de Gier, B. Nienhuis, \emph{J.
Phys. A:~Math. Gen.} \textbf{34} (2001) L265}

\bibitem{KQ} {\small C. Korff \emph{J. Phys. A: Math. Gen.} \textbf{36}
(2003) 5229}

\bibitem{KQ2} {\small C. Korff \emph{J. Phys. A: Math. Gen}. \textbf{37}
(2004) 385}

\bibitem{KQ3} {\small C. Korff \emph{J. Phys. A: Math. Gen}. \textbf{37}
(2004) 7227}

\bibitem{KQ4} {\small C. Korff \emph{Auxiliary matrices on both sides of the
equator}. math-ph/0408023 (to appear in J. Phys. A: Math. Gen. 37, December
issue)}

\bibitem{KQ5} {\small C. Korff \emph{Solving Baxter's TQ equation via
representation theory.} Proceedings "Non-Commutative Geometry and
Representation Theory in Mathematical Physics", July 2004, Karlstads
Universitet, Karlstad, Sweden; math-ph/0411034}

\bibitem{BS90} {\small V. V. Bazhanov and Yu. G.~Stroganov~\emph{J. Stat.
Phys.}\ \textbf{59} (1990) 799}

\bibitem{AF97} {\small A. Antonov and B. Feigin \emph{Phys. Lett. B} \textbf{%
392} (1997) 115}

\bibitem{RW02} {\small M. Rossi and R.~Weston~ \emph{J. Phys. A: Math. Gen}. 
\textbf{35} (2002) 10015.}

\bibitem{BLZ99} {\small V.~Bazhanov , S. Lukyanov and A. Zamolodchikov, 
\emph{Comm. Math. Phys}. \textbf{200} (1999) 297}

\bibitem{ODEIM} {\small P. Dorey, R. Tateo \emph{J. Phys. A: Math. Gen. }%
\textbf{32}\emph{\ }(1999) L419;}

{\small V. Bazhanov, S.~Lukyanov and A. Zamolodchikov \emph{J. Stat. Phys.} 
\textbf{102} (2001) 567; \emph{Adv. Theor. Math. Phys.} \textbf{7} (2003) 711%
}

\bibitem{CPbook} {\small V. Chari and A.~Pressley \emph{A Guide to Quantum
Groups}, Cambridge Univ. Press, 1994}

\bibitem{Drin} {\small V.G.~Drinfel'd \emph{Quantum groups}. In: A.M.
Gleason (ed.) Proceedings of the 1986 International Congress of Mathematics,
Berkeley, pp.798-820. Providence, RI: AMS 1987}

\bibitem{Jimbo} {\small M.~Jimbo Lett. Math. Phys.\ 10 (1985) 63}

\bibitem{dCKP} {\small C.~de Concini and V.~Kac. Representations of Quantum
Groups at Roots of 1. In \emph{Operator algebras, unitary representations,
enveloping algebras, and invariant theory, }eds. A. Connes et al., Progress
in Mathematical Physics 92, Birkh\"{a}user, 1990, pp 471; C.~De Concini,~V.
Kac, C. Procesi J. Amer. Math. Soc. \textbf{5} (1992) 151}

\bibitem{CP97} {\small V. Chari and A. Pressley \emph{Representation Theory} 
\textbf{1} (1997) 280}

\bibitem{RA} {\small P. Roche and D. Arnaudon \emph{Lett. Math. Phys}. 
\textbf{17} (1989) 295}

\bibitem{Ktw} {\small C.~Korff, \emph{J. Phys. A: Math. Gen.} \textbf{37}
(2004) 1681}
\end{thebibliography}
\end{document}